\documentclass[aps,prl,twocolumn,showpacs,superscriptaddress,preprintnumbers,amsmath,amssymb]{revtex4-1}

\usepackage{graphicx}
\usepackage{dcolumn}
\usepackage{bm}
\usepackage[amssymb]{SIunits}
\usepackage{color}
\usepackage{textcomp}
\usepackage{braket}

\newcommand{\siv}{SiV\textsuperscript{-} }

\begin{document}

\preprint{APS/123-QED}

\title{All-optical control of the silicon-vacancy spin in diamond at millikelvin temperatures}

\author{Jonas N. Becker}
\affiliation{Naturwissenschaftlich-Technische Fakult\"at, Fachbereich Physik, Universit\"at des Saarlandes, Campus E2.6, 66123 Saarbr\"ucken, Germany}

\author{Benjamin Pingault}
\affiliation{Cavendish Laboratory, University of Cambridge, JJ Thomson Ave, Cambridge CB3 0HE, UK}

\author{David Gro\ss{}}
\affiliation{Naturwissenschaftlich-Technische Fakult\"at, Fachbereich Physik, Universit\"at des Saarlandes, Campus E2.6, 66123 Saarbr\"ucken, Germany}

\author{Mustafa  G\"{u}ndo\u{g}an}
\affiliation{Cavendish Laboratory, University of Cambridge, JJ Thomson Ave, Cambridge CB3 0HE, UK}

\author{Nadezhda Kukharchyk}
\affiliation{Naturwissenschaftlich-Technische Fakult\"at, Fachbereich Physik, Universit\"at des Saarlandes, Campus E2.6, 66123 Saarbr\"ucken, Germany}

\author{Matthew Markham}
\affiliation{Global Innovation Centre, Element Six Limited, Global Innovation Centre, Fermi Avenue, Harwell Oxford, Didcot, Oxfordshire OX11 0QR, UK}

\author{Andrew Edmonds}
\affiliation{Global Innovation Centre, Element Six Limited, Global Innovation Centre, Fermi Avenue, Harwell Oxford, Didcot, Oxfordshire OX11 0QR, UK}

\author{Mete Atat\"{u}re}
\thanks{Electronic address: ma424@cam.ac.uk}
\affiliation{Cavendish Laboratory, University of Cambridge, JJ Thomson Ave, Cambridge CB3 0HE, UK}
\email{ma424@cam.ac.uk}

\author{Pavel Bushev}
\thanks{Electronic address: pavel.bushev@physik.uni-saarland.de}
\affiliation{Naturwissenschaftlich-Technische Fakult\"at, Fachbereich Physik, Universit\"at des Saarlandes, Campus E2.6, 66123 Saarbr\"ucken, Germany}
\email{@physik.uni-saarland.de}

\author{Christoph Becher}
\thanks{Electronic address: christoph.becher@physik.uni-saarland.de}
\affiliation{Naturwissenschaftlich-Technische Fakult\"at, Fachbereich Physik, Universit\"at des Saarlandes, Campus E2.6, 66123 Saarbr\"ucken, Germany}
\email{christoph.becher@physik.uni-saarland.de}

\date{\today}

\begin{abstract}
The silicon-vacancy center in diamond offers attractive opportunities in quantum photonics due to its favorable optical properties and optically addressable electronic spin. Here, we combine both to achieve all-optical coherent control of its spin states. We utilize this method to explore spin dephasing effects in an impurity-rich sample beyond the limit of phonon-induced decoherence: Employing Ramsey and Hahn-echo techniques at 12\,mK base temperature we identify resonant coupling to a substitutional nitrogen spin bath as the limiting decoherence source for the electron spin.
\end{abstract}

\maketitle

Colour centers in diamond are atomic-sized optically active impurities incorporated in the diamond lattice. During the past decade, they have been established as versatile tools for solid-state quantum information processing (QIP), quantum-enhanced sensing as well as biolabelling applications \cite{acosta2013}. While the nitrogen-vacancy center (NV\textsuperscript{-}) so far largely dominates this field of research, its poor optical properties, such as intense phonon sidebands, highly limit its applicability in QIP and its future success depends on the development of efficient photonic interfaces \cite{dam2017}. In contrast, the negatively charged silicon-vacancy center (SiV\textsuperscript{-}) in diamond offers remarkable optical properties \cite{hepp2014,rogers20143,jantzen2016}, among which fluorescence concentrated into the zero-phonon line at 737\,nm and low inhomogeneous broadening, making it a promising building block for photonic quantum networks \cite{obrien2009,kimble2008}.
\begin{figure}[t]
	\centering
	\includegraphics[width=0.45\textwidth]{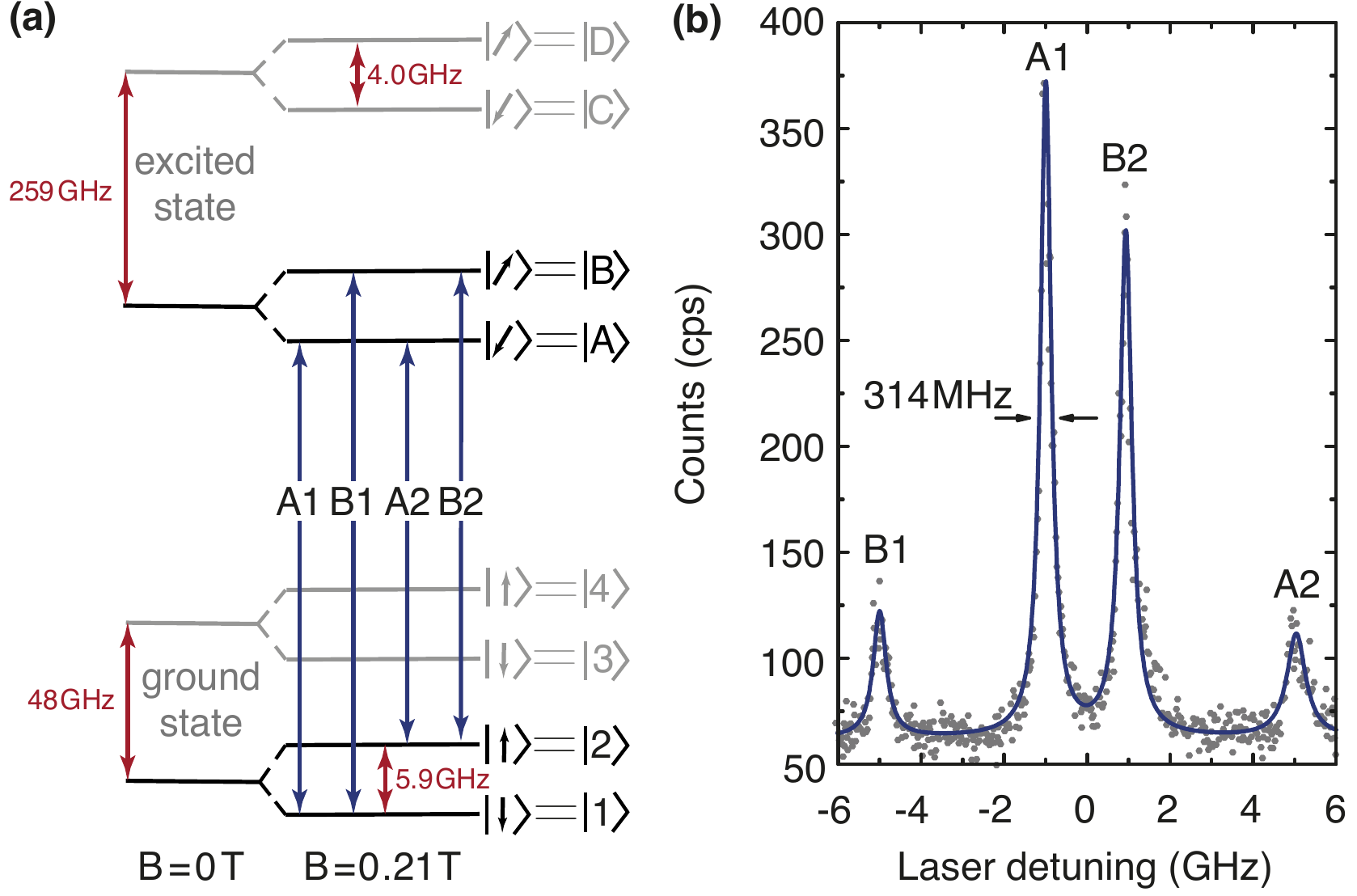}\\
	\caption{(color online) (a) Electronic level scheme of the \siv at B=0\,T (left) and B=0.21\,T (right). Optical transitions are indicated by blue arrows, splittings by red arrows. (b) Photoluminescence excitation spectrum of the transitions between the lowest ground and excited state spin doublets at 3.7\,K.}
		\label{fig:Fig1}
\end{figure}
The energy levels of the \siv can also be used to store and process information and thus their interfacing with photons is an enabling step for the implementation of QIP \cite{sipahigil2016}. Towards this goal, ultrafast all-optical control of the orbital degree of freedom of the \siv has been achieved recently \cite{becker2016,zhou2017}. However, the short coherence times associated with these orbital levels limits their usability. On the contrary, the spin S=1/2 of the \siv offers longer coherence times \cite{pingault2014,rogers2014,pingault2017} and thus presents itself as a desirable qubit. While microwave-based coherent control of the spin has been reported \cite{pingault2017}, all-optical control gives access to ultra-fast spin manipulation, allowing for thousands of spin rotations, even in the presence of fast decoherence processes as they are common in solid-state matrices. Furthermore, the low power typically required for all-optical coherent control schemes circumvents heating effects often present with microwave control. In this Letter, we demonstrate all-optical control of the \siv electronic spin. We then make use of this control to probe and analyse spin decoherence of the center at millikelvin temperatures.\\
In experiments performed at 4\,K, the spin dephasing time is limited to approximately 100\,ns \cite{pingault2014,rogers2014,pingault2017} with spin decoherence being dominated by transitions among orbital ground-state branches mediated by phonons at appromimately 50\,GHz \cite{jahnke2015,pingault2017}. To circumscribe this limitation, several methods have been proposed, such as reducing the phonon density of states using nanostructures or applying strain to the \siv to split the ground state levels apart \cite{sohn2017}. A more straightforward approach consists in operating at lower temperatures to reduce the 50-GHz phonon population to a negligible level, which we realize here using a dilution refrigerator reaching a measured base temperature of 12\,mK. The suppression of phonon-induced spin decoherence allows to analyse the spin dephasing processes due to coupling to the environment of a sample with a considerable spin impurity level. We find that the dominant mechanism arises from resonant interaction of the \siv spin with the spin bath formed by substitutional nitrogen atoms.\\
We investigate a single \siv center hosted in a (111)-oriented HPHT IIa bulk diamond containing a large concentration of substitutional nitrogen impurities, a common residual impurity in this type of diamond \cite{sumiya1996}. This is the same emitter used in \cite{becker2016} and the spectroscopic characterization of this emitter at zero magnetic field can be found therein. The diamond sample is cooled down to millikelvin temperatures in a dilution refrigerator with free-space optical access using a home-built confocal setup \cite{Kukharchyk2017}. The setup is capable of operating at millikelvin temperatures and consists of a copper cage carrying a titanium slip-stick positioner stack with a copper sample holder containing a SmCo permanent magnet and a numerical aperture of 0.9 achromatic microscope objective \cite{suppl}. In the absence of external magnetic field, the energy levels of the SiV\textsuperscript{-}, displayed in Fig.\,\ref{fig:Fig1}(a), consists of two-fold orbitally split ground and excited states, with a splitting of 48\,GHz between the two branches of the ground state. We lift the degeneracy of the electronic spin S=1/2 by applying a magnetic field of 0.21\,T at an angle of 70.5\,$^{\circ}$ with the \siv symmetry axis, resulting in the Zeeman splitting of the orbital branches, as shown in Fig.\,\ref{fig:Fig1}(a) with the spin projection indicated for each level. At such an angle between the applied magnetic field and the \siv axis, optical transitions between all levels are allowed due to a difference in ground and excited state quantization axes \cite{hepp2014,pingault2014}. This is exemplified by scanning the frequency of a Ti:Sapphire laser to excite resonantly the transitions labelled by blue arrows in Fig.\,\ref{fig:Fig1}(a) and detecting the resulting fluorescence on the phonon sideband of the \siv (photoluminescence excitation spectrum, PLE), as shown in Fig.\,\ref{fig:Fig1}(b). During cooldown to 12\,mK a broadening of the linewidths in the PLE can be observed, as the vanishing phonon-induced transitions between the ground states cause an increase in the effective spin-pumping rates and thus a shortening of the effective ground state relaxation times. This can be remedied by adding a second laser to counteract spin-initialization and this effect, known as light-narrowing, is a clear sign for a change in the internal dynamics of the center \cite{suppl}.\\
Next, we use the laser carrier and sideband generated by an electro-optical modulator (EOM) to excite resonantly transitions A1 and A2 simultaneously, which form a $\Lambda$-system between two ground state levels $\ket{1}$ and $\ket{2}$ of opposite spin, as seen in Fig.\,\ref{fig:Fig1}(a). Moreover, we use acousto-optical modulators (AOMs) to generate 10-ns pulses resonant with each transition, thus driving Raman transitions between $\ket{1}$ and $\ket{2}$. We measure the fluorescence arising from residual excitation into the excited state as a function of time during the laser pulses, as displayed in Fig.\,\ref{fig:Fig2}(inset).
\begin{figure}[t]
	\centering
	\includegraphics[width=0.45\textwidth]{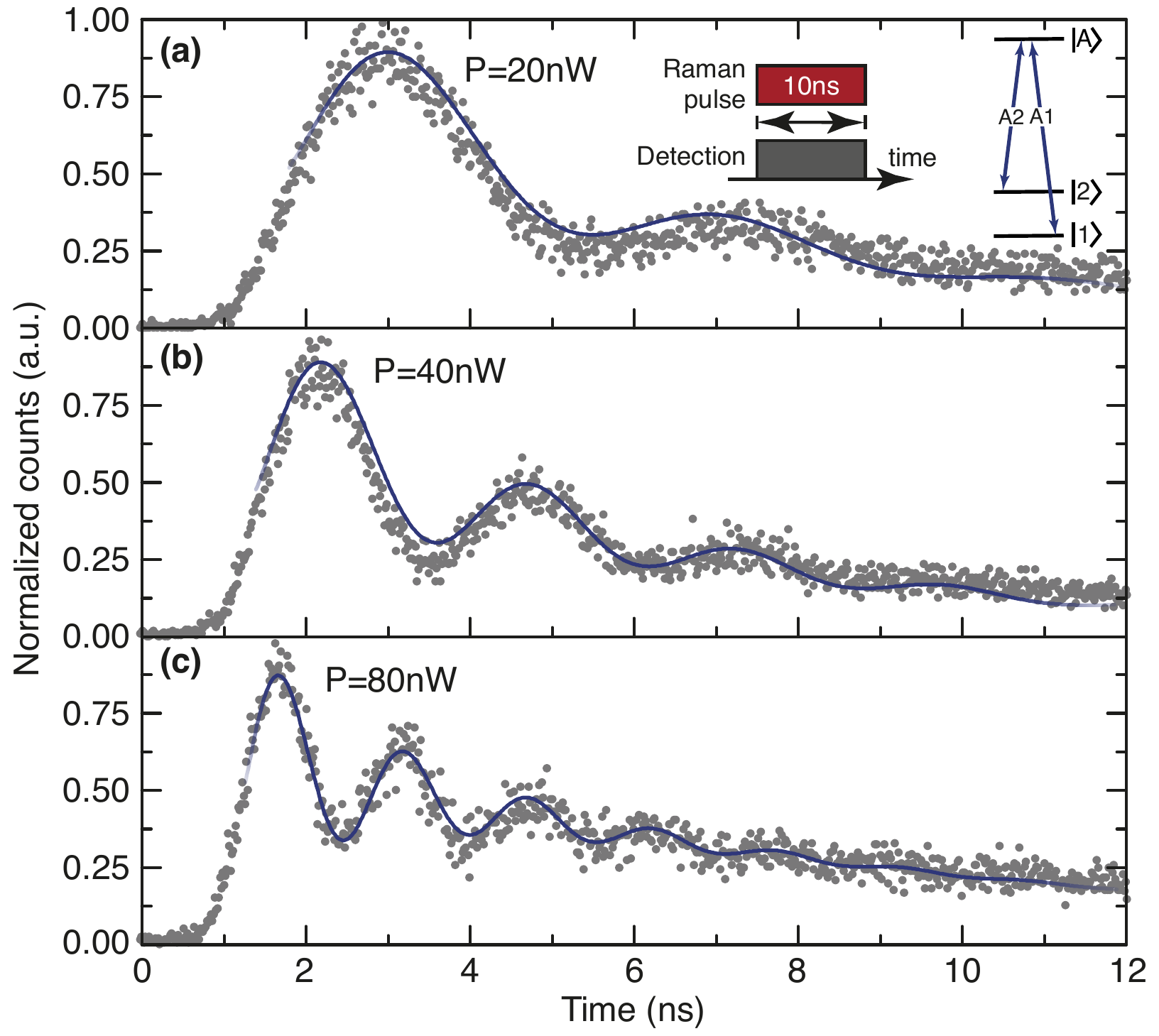}\\
	\caption{(color online) All-optical Rabi oscillations between the two lowest ground state spin sublevels $\ket{1}$ and $\ket{2}$ using a bi-chromatic Raman pulse at 12\,mK. As expected for two-photon Rabi oscillations, the Rabi frequencies scales linearly ($\Omega$(20\,nW)=1.54\,MHz, $\Omega$(40\,nW)=2.48\,MHz, $\Omega$(80\,nW)=4.13\,MHz) with power. Blue lines represent simulations using a four-level Bloch equation model discussed in \cite{pingault2014}.}
		\label{fig:Fig2}
\end{figure}
The fluorescence oscillates in time, indicating Raman-Rabi oscillations between the two ground state levels, as shown in Fig.\,\ref{fig:Fig2}(a), thus demonstrating all-optical control of the spin of the SiV\textsuperscript{-}. A four-level master equation model described in Ref.\,\cite{pingault2014} reproduces the Rabi oscillations very well, as shown by solid blue curves. The downward drift of the oscillations is due to a combination of the imperfect optical pulse shape featuring a slowly decaying falling edge and the spontaneous decay from the excited state. Both effects lead to an exponential drop-off of the mean fluorescence level. The combined effect has been taken into account by introducing a common exponential amplitude fit factor to the model which however leaves contrast and frequency of the Rabi oscillations unaltered. The experiment is repeated at several laser powers shown in Fig.\,\ref{fig:Fig2}(b,c) to confirm that the Rabi frequency scales linearly with laser power, as expected for two-photon transitions.\\
We make use of the all-optical spin control to explore spin coherence via Ramsey interferometry using two $\pi/2$ pulses separated by a variable delay $\tau$, as illustrated in Fig.\,\ref{fig:Fig3}(a,inset), for a temperature of 12\,mK.
\begin{figure}[t]
	\centering
	\includegraphics[width=0.45\textwidth]{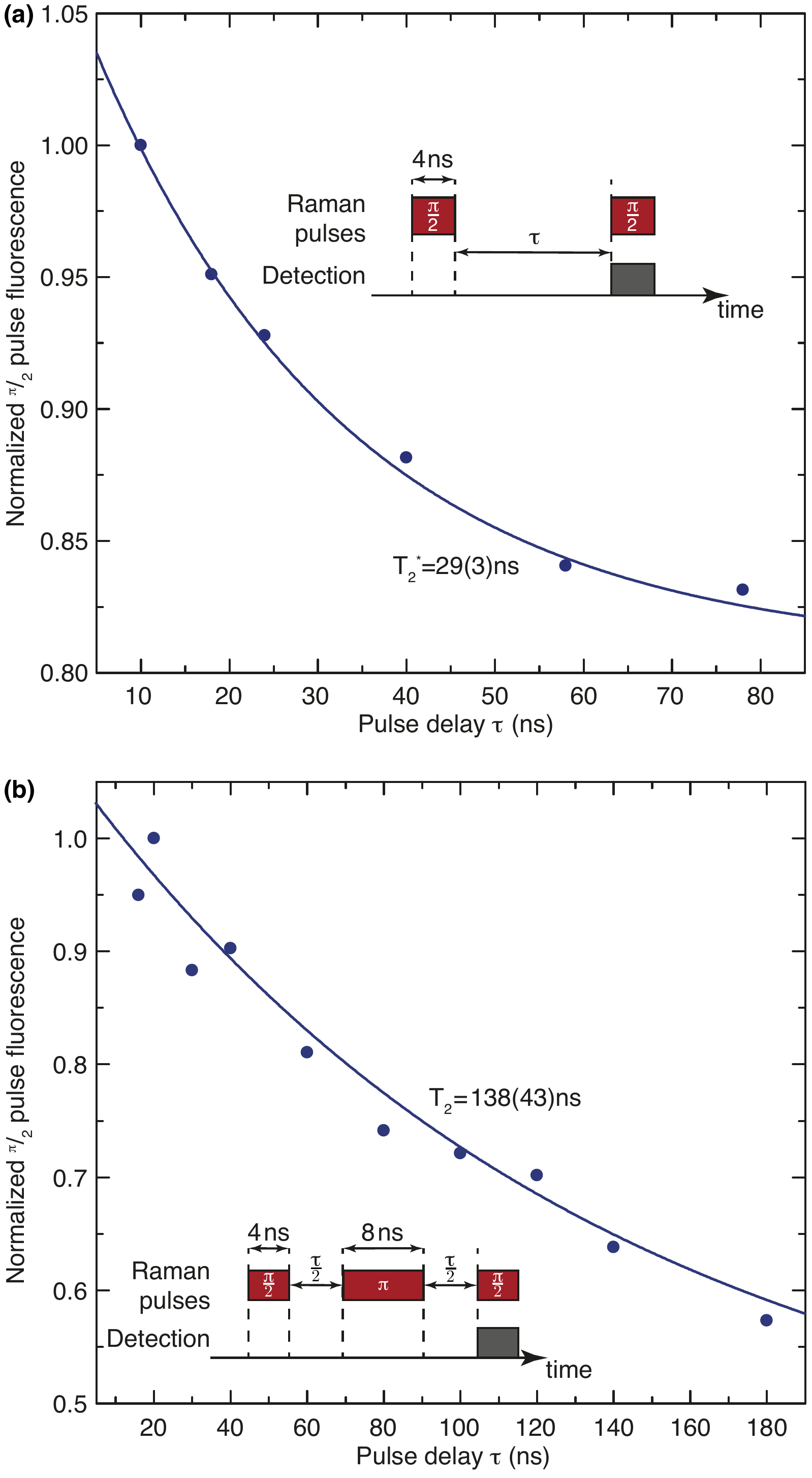}\\
	\caption{(color online) (a) Ramsey interference envelope measured at 12\,mK by applying two subsequent 4\,ns long $\frac{\pi}{2}$ Raman pulses (average power P=0.5\,nW) with a number of fixed delays and maximizing the fluorescence in the second pulse by fine-tuning their temporal spacing. (b) Spin echo measurement using a simple three pulse echo sequence with a single refocusing $\pi$ pulse (average power P=1\,nW). Blue lines in are simulations using the above-mentioned density matrix model with the power ratio of both Raman components as well as the ground state decoherence rate as free parameters.}
		\label{fig:Fig3}
\end{figure}
The spin population is read out by measuring the fluorescence emitted during the second $\pi/2$-pulse. The envelope of the interference pattern, displayed in Fig.\,\ref{fig:Fig3}(a) as blue dots, is seen to decrease exponentially with the delay $\tau$, yielding a spin dephasing time T$_2^*$=29(3)\,ns, in good agreement with the value extracted through an independent experiment using coherent population trapping \cite{suppl}.
\begin{figure*}[htb]
	\centering
	\includegraphics[width=1\textwidth]{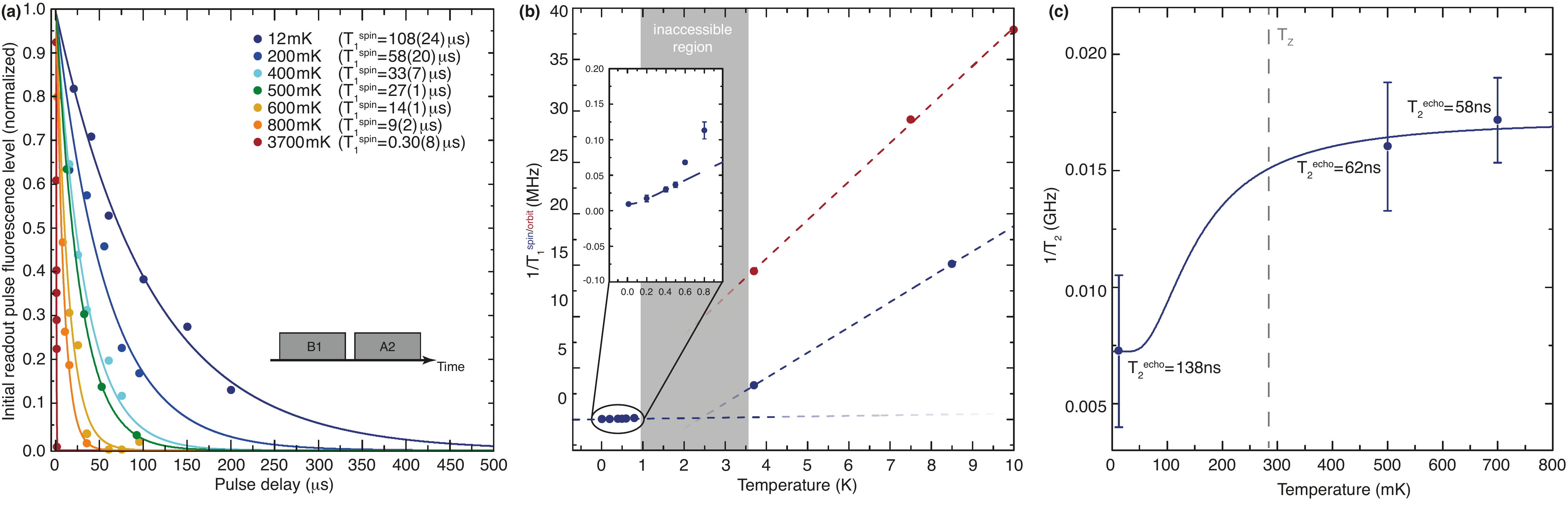}\\
	\caption{(color online) (a) Spin relaxation measured via a two-pulse optical pumping sequence on transitions B1 and A2 for different temperatures between 3.7\,K and 12\,mK and an average optical power of P=50\,nW. An exponential fit to the decay of the fluorescence signal yields the spin relaxation time T$_1^\text{spin}$. (b) Temperature-dependent spin (blue) and orbital relaxation rates (red) indicating single-phonon spin relaxation processes below 300\,mK and above 2.3\,K as well as non-linear multi-phonon processes in between. Insert shows zoom of the low temperature region. Further details can be found in the main text. The grey temperature region (1-3.7\,K) is unaccessible with the cryostat used in this study. (c) Temperature-dependent spin echo measurements according to the scheme in Fig.\,\ref{fig:Fig3}(b). Blue line corresponds to an analytical model including a residual decoherence rate at 0\,K as well as a temperature- and magnetic field-dependent decoherence rate describing field fluctuations generated by flip-flop processes in a spin bath (C=2$\pi\cdot$6.34\,kHz, $\Gamma_{res}$=2$\pi\cdot$1.15\,MHz, T$_Z$=280\,mK).}
	\label{fig:Fig4}
\end{figure*}
In order to explore the limits of the phase coherence, we implement a Hahn-echo sequence whereby we add a refocusing $\pi$-pulse between the two $\pi/2$-pulses of the Ramsey sequence, as depicted in Fig.\,\ref{fig:Fig3}(b,inset). Fig.\,\ref{fig:Fig3}(b) displays the decay of the envelope of the signal. Using an exponential fit, we obtain a spin echo time T$_{2,\text{echo}}$=138(43)\,ns at 12\,mK.\\
The above-discussed improvement by a factor 4.8 compared to T$_2^*$ owing to the refocusing $\pi$-pulse indicates a non-Markovian dephasing mechanism. It is also interesting to note that the spin dephasing time T$_2^*$ does not change considerably when cooling from 3.7\,K (T$_2^*$=20\,ns \cite{suppl}) to 12\,mK (T$_2^*$=29\,ns). To understand this unexpected scaling we in the following further investigate the contributions of different decoherence mechanisms. First, we measure the spin relaxation time T$_1^\text{spin}$ between $\ket{1}$ and $\ket{2}$ using optical pumping \cite{rogers2014}, showing an improvement by a factor of 300 upon cool-down, starting from T$_1^\text{spin}$=303(8)\,ns at 3.7\,K and reaching T$_1^\text{spin}$=108(24)\textmu s at 12\,mK (Fig.\,\ref{fig:Fig4}(a)). This result confirms the suppression of phonon-driven relaxation processes via cooling and sets an ultimate limit of T$_2^*$=2T$_1$=216\,\textmu s to the spin coherence time in the absence of any additional dephasing. The long relaxation time now allows for efficient state initialization: By pumping into the lower spin sublevel $\ket{1}$ via transition A2, a spin initialization fidelity of at least 99.93\%, limited by detector dark counts, has been achieved \cite{suppl}.\\
From temperature-dependent T$_1^\text{spin}$ measurements (Fig.\,\ref{fig:Fig4}(b)) we identify four different regimes: (i) for T$\leq$50\,mK a saturation indicating an effective temperature of the SiV of about 40\,mK \cite{abragam}; for 50\,mK$<$T$<$500\,mK a linear variation pointing at a direct spin relaxation process \cite{shrivastava1983}; (iii) for 500\,mK$<$T$<$2.3\,K a non-linear temperature dependence of T$_1^\text{spin}$ hinting at higher order phonon processes like Raman or Orbach processes \cite{shrivastava1983,bogani2008}. To further investigate the exact nature of this process additional measurements at intermediate temperatures are necessary but not feasible with the employed cryostat. (iv) For T$\geq$2.3\,K the relaxation rate then scales again linearly due to the dominant one-photon process involving excitation to the upper orbital ground state branch as discussed above. In this temperature regime, the spin relaxation T$_1^\text{spin}$ is however slower than the orbital relaxation T$_1^\text{orbit}$ (red dots in Fig.\,\ref{fig:Fig4}(b)). We note that the measured spin coherence times ranging from T$_2^*$=20\,ns at 3.7\,K to T$_2^*$=29\,ns at 12\,mK \cite{suppl} are not limited by any of these effects but have to be influenced by further decoherence processes.\\
To identify the source of residual spin decoherence we performed temperature dependent spin-echo measurements, displayed in Fig.\,\ref{fig:Fig4}(c). We attribute the spin decoherence to an interaction with the substitutional nitrogen (P1 center, S=1/2, g=2) spin bath typically present in HPHT diamond. This is supported by the observed temperature dependence which can be well-fitted by a model of the form
\begin{equation}
\frac{1}{\text{T}_{2,\text{echo}}}=\frac{\text{C}}{(1+\text{exp}(\frac{\text{T}_{\text{Z}}}{\text{T}}))(1+\text{exp}(-\frac{\text{T}_{\text{Z}}}{\text{T}}))}+\Gamma_{\text{res}}.
\end{equation}
This model contains two decoherence processes associated with the spin bath \cite{kutter1995,takahashi2008}: (i) the first term represents decoherence caused by magnetic field fluctuations created by energy-conserving spin flip-flop processes between bath spins. The rephasing of this noise is what causes the improvement of T$_{2,\text{echo}}$ compared to T$_2^*$. Since the frequency of this noise increases with the flip-flop rate at higher temperatures or lower magnetic fields, this noise becomes more and more Markovian and hence will be rephased less well for increasing temperatures. (ii) The second term, $\Gamma_\text{res}$, corresponds to a residual decoherence rate at 0\,K, caused by a resonant dipolar coupling of the \siv to the surrounding bath spins. This occurs since the SiV\textsuperscript{-}, like the bath spins, is a S=1/2 spin system with g$\approx$2 and therefore experiences the same Zeeman splitting as the P1 bath at all magnetic fields. This process is entirely Markovian in nature and cannot be counteracted by spin-echo or dynamic decoupling sequences. This effect can also be observed in the NV\textsuperscript{-} for specific magnetic field strengths at which its $\ket{m_s=−1}\to\ket{m_s=0}$ transition is tuned into resonance with the bath, resulting in a strongly enhanced decoherence by several orders of magnitude \cite{hanson2006}. The zero-field splitting of the NV\textsuperscript{-} results in reaching resonance with the P1 bath only at 0.054\,T and thus operating away from this condition results in much longer coherence times, even in nitrogen-rich diamonds. For the SiV\textsuperscript{-}, we expect this coupling to decrease significantly by improving sample purities, making longer coherence times feasible. For quantum technology applications, the \siv spin qubit coherence will then be ultimately limited by the much weaker dephasing due to a residual nuclear spin bath (e.g. $^{13}$C) or the T$_1$ limits as explored above.\\
In the limit of strong coupling to the bath spins, we can estimate an average spin bath density according to Hu et al.\cite{hu1974} of 3.8\,ppm \cite{suppl}, about one to two orders of magnitude higher than what is expected for high-quality type IIa diamond \cite{sumiya1996}. The spin bath density might however vary strongly across the sample \cite{babich2006}. These local variations also account for the severe difference in coherence times observed for (bright) \siv centers in different locations of the sample (e.g. T$_2^*=$45\,ns for the emitter investigated by Pingault et al.\cite{pingault2014} in the same sample). By selecting exceptionally bright \siv centers for the investigations we might have undeliberately targeted high density regions of the sample: The \siv centers in the HPHT IIa material used here overall show an emission enhanced by one to two orders of magnitude compared to centers in e.g. chemical-vapor deposited IIa diamond containing considerably less nitrogen. This as well as additional diamond growth experiments \cite{singh2013} suggests a correlation between brightness and stability of the \siv emission as well as the abundance of electron donors such as P1. Furthermore, another \siv in identical type IIa HPHT material which has recently been used in \cite{pingault2017}, shows practically identical T$_1^\text{orbit}$ times to the ones obtained here but a spin coherence nicely matching T$_2^*$=2T$_1^\text{orbit}$=134(4)\,ns at 3.5\,K. This suggests that this \siv is located in a relatively pure region of the sample and, in accordance with the above-mentioned hypothesis, this \siv displayed significantly lower count rates under resonant excitation, indicating potential charge state issues. Thus we conclude that electron donor impurities like the P1 on the one hand can severely affect the spin coherence time of the \siv but might on the other hand be responsible for the significant fluorescence enhancement observed for \siv centers examined in type IIa HPHT material compared to centers in the much purer electronic grade CVD diamond, enabling the all-optical spin control presented here.\\
In conclusion, we here have demonstrated all-optical coherent control of the \siv electron spin at millikelvin temperatures, laying the foundation for ultrafast spin qubit manipulation in future experiments. We utilized this technique to measure the coherence properties of single \siv centers in an impurity-rich bulk diamond at temperatures as low as 12\,mK. The suppression of phonon population leads to an improvement in spin relaxation times by a factor of 300. The spin coherence, however, is only weakly affected by the phonon suppression: we find T$_2^*$=29\,ns at 12 mK and a simple spin echo technique improves the coherence time by a factor of 5, partially remedying slow magnetic field noise created by an electron spin bath of the sample. While this noise contribution can be further suppressed in future experiments by employing more complex dynamic decoupling pulse sequences \cite{suter2016} an ultimate limit is set by a resonant coupling to the nitrogen bath which cannot be rephased or decoupled and can only be improved in samples containing less nitrogen. This emphasizes the need for a detailed diamond growth study to fabricate \siv samples combining the excellent optical properties of the \siv with good spin properties paving the way towards efficient spin photon interfaces \cite{togan2010}, especially in combination with integrated diamond nanophotonic platforms \cite{sipahigil2016}. Moreover, the confocal setup operating at millikelvin temperatures will be a highly valuable tool beyond the scope of this work to probe the physics of e.g. single rare earth ions or molecules in the still largely unexplored millikelvin regime.

\begin{acknowledgments}
This research has been partially funded by the European Community's Seventh Framework Programme (FP7/2007-2013) under Grant agreement No. 611143 (DIADEMS). Mete Atat\"ure gratefully acknowledges financial support by the European Research Council ERC Consolidator Grant Agreement No. 617985 and the EPSRC National Quantum Technologies Programme NQIT EP/M013243/1. Benjamin Pingault thanks Wolfson College (University of Cambridge) for support through a research fellowship. Ion implantation was performed at and supported by RUBION, the central unit of the Ruhr-Universit\"at Bochum. We thank D. Rogalla for the implantation, C. Pauly for the fabrication of solid immersion lenses. Moreover, we thank Camille Stavrakas for her help during early stages of the experiments and Johannes G\"orlitz for help with the laser systems. We also thank Elke Neu, Mark Newton, Ben Green and Phil Diggle for helpful discussions of spin bath physics in diamond and Dirk Englund and Matt Trusheim for providing additional samples.
\end{acknowledgments}


%

\end{document}